# Revisiting dynamics of Sun center relative to barycenter of Solar system

***or***

# Can we move towards stars using Solar self-resulting photo-gravitational force?

**Sergey Ershkov***, Affiliation[1]: Plekhanov Russian University of Economics,

Scopus number 60030998, e-mail: sergej-ershkov@yandex.ru

Affiliation[2]: Sternberg Astronomical Institute, M.V. Lomonosov's Moscow State University, 13 Universitetskij prospect, Moscow 119992, Russia,

**Dmytro Leshchenko**, Odessa State Academy of Civil Engineering

and Architecture, Odessa, Ukraine, e-mail: leshchenko_d@ukr.net

**Abstract**

We introduce here in the current research the revisiting of approach to the dynamics of Sun center relative to barycenter of Solar system by using self-resulting photo-gravitational force of the Sun as the main reason of such motion. In case of slowly moving in the direction outwards with respect to the initial position of barycenter of Solar system (together with the current position of Solar system barycenter, of course) with average established velocity not less than 1050 Km/day, we should especially note that hierarchical configuration of Solar system will be preferably the same during this motion. As the main findings, we have suggested algorithm how to move towards stars using Solar self-resulting photo-gravitational force. The obvious physically reasonable assumption is that the Solar system will have been increasing its size during the evolution in a future (due to losses of the total angular momentum taking into account the tidal phenomena).

## 1. Introduction, dynamics of Sun center relative to barycenter of Solar system.

A lot of meaningful attempts have been made during last 60 years in the field of Celestial mechanics for theoretical describing the dynamics of solar center relative to barycenter of Solar system [1-6]. The main forces causing dynamical response of Sun during the aforementioned process yield exceptionally its *planar* nonlinear motion within the solar system's invariable plane [7] in close vicinity of barycenter of all the planets and Sun, Fig.1 (e.g., work [1] treats the Sun as a regular body in the solar system), taking into account that, *videlicet*, such motion takes place mainly under the action of mutual gravitational forces. Meanwhile, center of Sun is moving around the aforementioned barycenter on quasi-periodic trajectory (less than 2.19 $R_{Sun}$ from barycenter [1], where $R_{Sun}$ is the radius of Sun).

In work [3] they use also the relativistic definition of the barycenter, take into account not only Newtonian forces between major planets, but also relativistic terms, the non-uniform gravitational potential of the Sun (J2 term only), the Earth and Moon (main harmonics), the influence of asteroids and dwarf planets, etc.

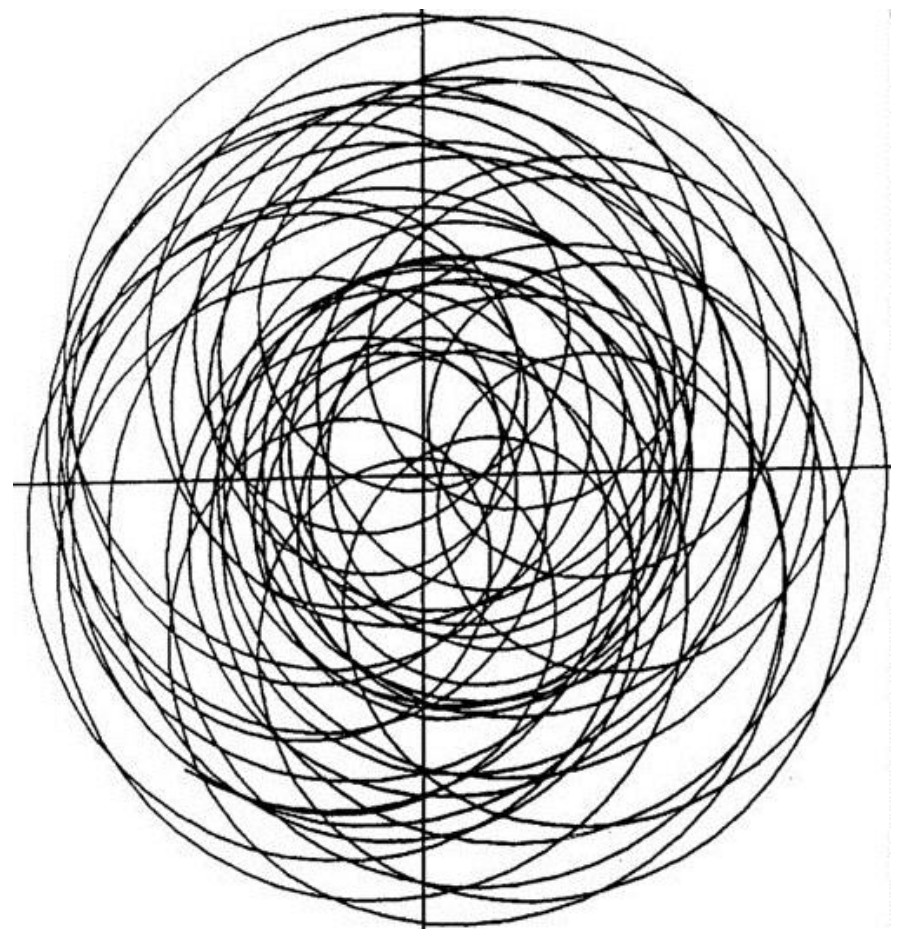

Fig.1. Schematically presented the oscillating of *planar* Sun center motion relative to barycenter of Solar system (including the Sun). The orbit of centre of Sun around barycenter of solar system (in units of $10^{-3}$ AU, in astronomical units; center of Sun

is moving within borders less than 2.19 $R_{Sun}$ from barycenter [1] or max. 0.01 AU), for details see [5].

The main idea of the current research is to revisit the dynamics of Sun center relative to barycenter of Solar system using self-resulting photo-gravitational force of the Sun. Indeed, translational motion of e.g. asteroids is known to be under the influence of effect of non-gravitational nature (Yarkovsky effect [8]). The Yarkovsky effect is the force acting on a rotating body in space caused by the anisotropic re-emission of thermal photons, which carry momentum [9-10]. Such a force is produced by the way an asteroid absorbs energy from the sun and re-radiates it into space as heat by anisotropic way.

Taking into consideration that there exists a disbalance of momentum when asteroid at first absorbs the light radiating from the sun (thenafterwards asteroid re-radiates the heat), we can assume that the similar effect exists to be applied in a large space-scale to the Sun itself whereas such effect should be provided by the self-resulting photo-gravitational force over all the surface of Sun which radiates light and energy fluxes into space obviously by anisotropical way. Such a disbalance is caused by the rotation of a Sun during period of its full turnover (e.g., there is effect of differential rotation in Solar convection zone [11]) as well as by the anisotropic cooling of its surface and inner layers; the processes above depend on anisotropic heat transfer in the inner layers of Solar convection zone [10].

During the years such disbalance forms a negligible, but important additional acceleration for the center of Sun due to aforedescribed effect. It is worthnoting that the resulting photo-gravitational force may play a crucial role in particular problems of Celestial Mechanics, e.g. at sudden cataclysmic explosion of a massive star (supernova) where photo-gravitational force is known to be much more exceeding the Newtonian force associated with central gravity field.

In the current research, we will restrict ourselves in presenting a new dynamical approach for explaining the dynamics of Sun center relative to barycenter of Solar system. As for the complete introduction to the problem under consideration, we recommend seminal works [1-4], where a significant historical retrospecting has been made as well as all the difficulties for such a kind of solar center motion are considered in details.

## 2. **Estimation of solar resultant (net) photo-gravitational force by its resulting contribution to the motion of Sun, in kinematic way.**

The *kinematic* ansatz means that instead of considering the dynamical causes [8] for the motion of Sun center relative to barycenter of Solar system, we should investigate its kinematic influence on elements of orbit, Figs. 2-3.

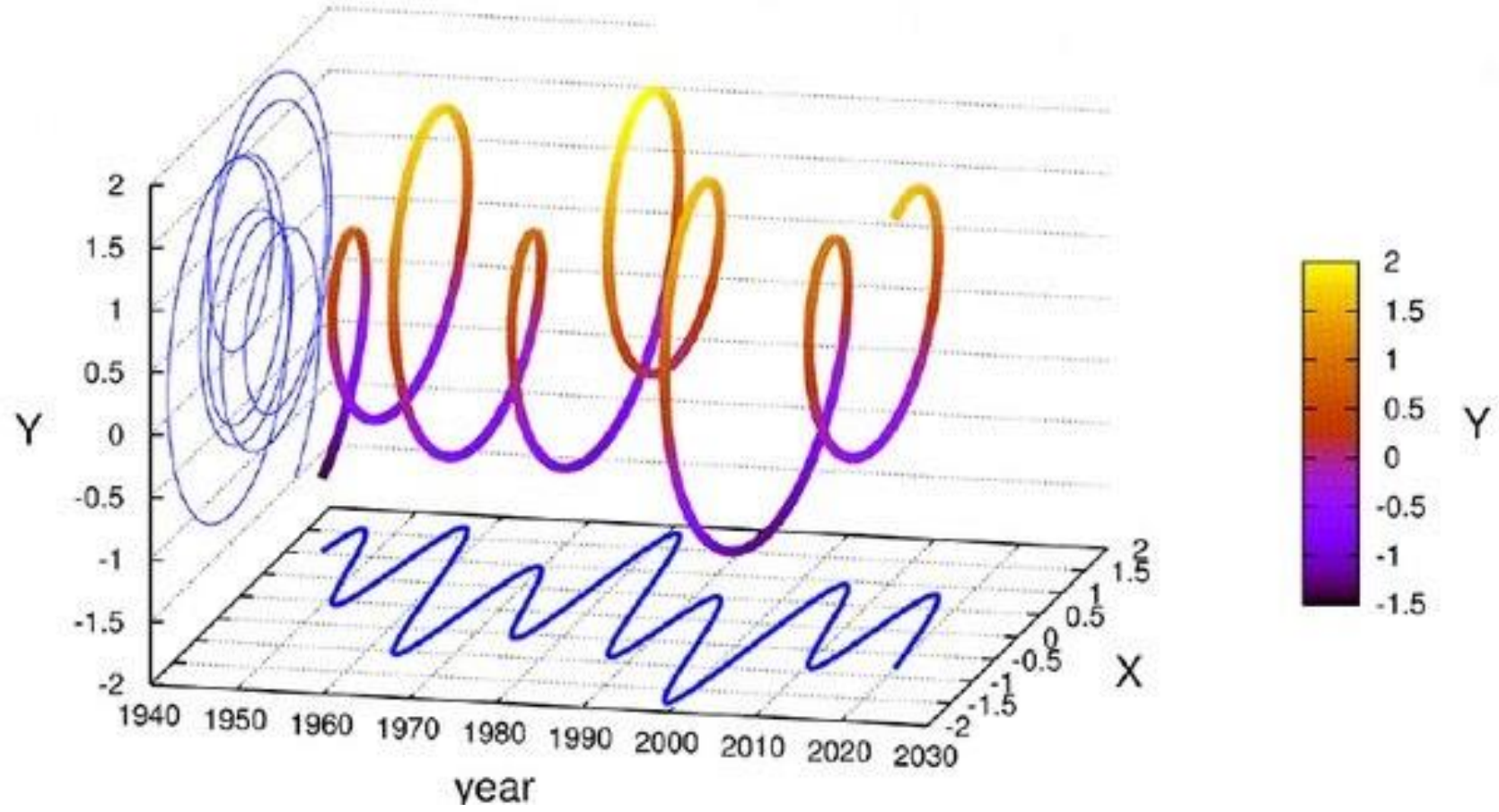

Fig.2. Schematically presented the kinematic resulting path of Sun center planar motion relative to barycenter of Solar system. Here abscissa-axis X and ordinate-axis Y correspond to axes of cartesian coordinate system with its origin located in the barycenter of solar system (for further details, scale, etc. see Fig.1).

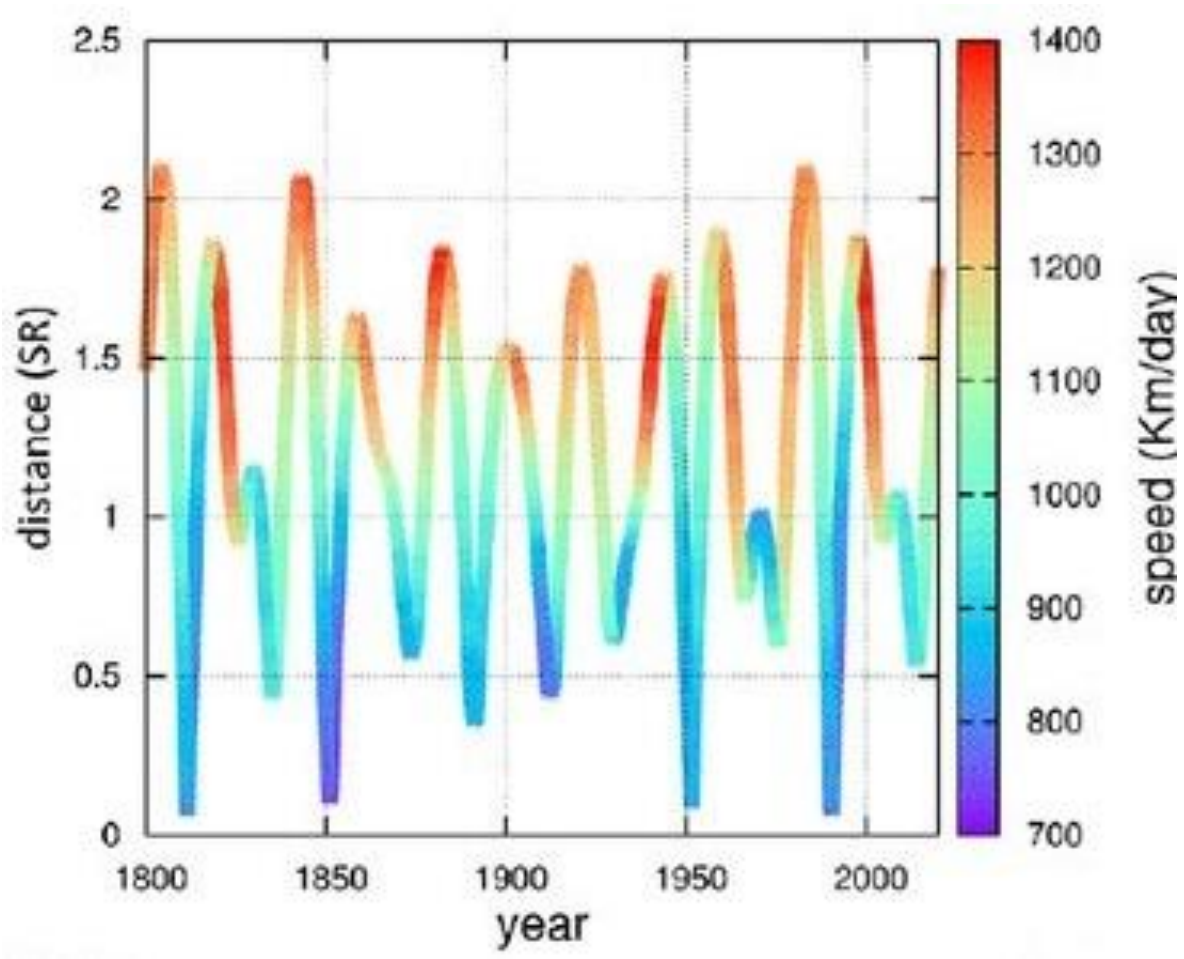

Fig.3. Schematically presented the *velocity dependence* on the resulting path of Sun center *planar* motion relative to barycenter of Solar system (for further details see Figs.1-2).

Let us estimate the solar resultant (net) photo-gravitational force by its resulting contribution to the motion of Sun, in kinematic way. Namely, we can make a certain conclusion from Figs. 2-3 that average meaning of the kinematic resulting distance of Sun center motion (relative to barycenter of Solar system) is circa 1.25 radius of Sun, whereas the average value of its velocity is circa 1050 Km/day. It means that if we succeed in straightening the curve of the resulting path of Sun center *planar* motion relative to barycenter of Solar system, Sun would may slowly move in the direction outwards with respect to the initial position of barycenter of Solar system (together with the current position of Solar system barycenter, of course) with average velocity not less than 1050 Km/day, ~44 Km/hour or 12 m/s. Such level of velocity seems not to be sufficient to overcome the distance between Sun and closest star (similar to Sun), Alpha Centauri, but nevertheless. It will take from us circa 102,000 years to go there successfully, if we obtain scientific technology which would help us to control the motion of Sun center motion.

We can add to the main conclusion in [2] (where was stated that the motion of planets influences solar activity) that the dynamics of Sun center relative to barycenter of Solar system mainly is governed by the self-resulting photo-gravitational force of the Sun (which should be calculated by integration over all the radiating surface of Sun as the net result of total solar activity coming from the Solar convection zone as well as from inner quasi-rigid core of the Sun).

## 3. Discussion & conclusion.

We introduce here in the current research the revisiting of approach to the dynamics of Sun center relative to barycenter of Solar system by using self-resulting photo-gravitational force of the Sun as the main reason of such motion.

As reported in [7], the total angular momentum of the Solar system is preferably conserved (we should note that tidal phenomena in Solar system are allowing the transfer of a infinitesimal amount of angular momentum from axial rotations to orbital revolutions due to tidal friction, and vice-versa). So, even in case of slowly moving in the direction outwards with respect to the initial position of barycenter of Solar system (together with the current position of Solar system barycenter, of course) with average velocity not less than 1050 Km/day, we should especially note that hierarchical configuration of Solar system will be preferably the same during this motion.

As the main findings, we have suggested algorithm how to move towards stars using Solar self-resulting photo-gravitational force. The obvious physically reasonable assumption is that the Solar system will have been increasing its size during the evolution in a future (due to losses of the total angular momentum taking into account the tidal phenomena), albeit we can not exclude the shrinking-scenario.

Referring to the rationale for the hypothesis how to suggest the way of straightening the curve of the resulting path of Sun center *planar* motion relative to barycenter of Solar system, we can formulate the main suggestion as follows (and its significance in the field of future practical using in Astronautica science field within the limits of what is necessary to indicate a realistic actual problem to which the results of the investigations can be applied): the local surface of Solar convection zone may be slightly changed by a series of explosion using nuclear weapon. Such deviating should be made very accurately (starting from negligible dozes of nuclear weapon exlosion), until we get a sufficiently long-term local distortion of the surface of Solar convection zone as the result of explosion by nuclear weapon. This will cause the local disbalance in self-resulting photo-gravitational force (driving Sun center motion relative to barycenter of Solar system) which will be caused by anisotropic radiating light and energy fluxes into space over all the surface of Sun during period of its full turnover, ~ circa 25 days.

The last but not least, it is worth to discuss now that the travelling to stars (without needs of leaving the Earth) together with Sun and all the solar system appears to be most comfortable technology as suggested above (instead of launching the spacecrafts and rockets), this could be planned, having understood in details and realized step-by-step insofar as clear and possible. It should be especially outlined that such way of travelling to star would be preferable to all the Earth's living creatures (not only human beings), i.e. those who would have been participants to form new branch of our civilization not only in the solar system (but in the vicinity of Alpha Centauri as well). As for the possibile physical mechanism explaining the possibility of Solar system motion itself, such natural algorithm appears to work similar to the phenomenon of self-resulting photo-gravitational force by its resulting contribution to the motion of Sun which should be, nevertheless, realized by avoiding the moving in its mean-motion *spirally* outwards with respect to the initial position of solar system barycenter during the centuries. Because straightening the curve of the resulting path of Sun center *planar* motion relative to initial position of barycenter of Solar system is, obviously, preferable. We should develop siginificantly the level of scientific knowledge and technology to drive such effect in automatical regime with minimal deviations on Sun center path to the chosen aim in the outer space (e.g., Alpha Centauri or Proxima Centauri).

All the discussed steps of the aforementioned approach are not fantastic, but scientifically reasonable stages of entire technology of using the Earth as giant Space Ark, in the same time suggesting the Sun as self-propelling Prime Mover (with help of gravitational towing of Earth, together with solar system, by the Sun).

Finally, achieving of the sufficient accuracy in the operating, targeting and tracking during the desired local distortion of surface of Solar convection zone by a series of explosion using nuclear weapon will eventually require the unifying all the strategic potentials of nuclear weapon for leading countries such as USA, China and Russia; it is also worthnoting that agreement within United Nations (an intergovernmental organization aiming to maintain international peace and security, develop friendly relations among nations) will to be required to approve such meaningful project.

Also, the remarkable articles [12-18] should be cited which concern the problem under consideration (meanwhile, in [13] author was absolutely right when stated that "The orbital motion of the Sun has been linked with solar variability, but the underlying physics remains unknown"). In [14] authors proposed constructive criticism or critical analysis of a hypothesis [15] of the planetary tidal influence on solar activity. Nevertheless, according to our opinion, authors were also absolutely right in [15] when stated that "correlations between direct solar activity indices and planetary configurations have been reported on many occasions. Since no successful physical mechanism was suggested to explain these correlations, the possible link between planetary motion and solar activity has been largely ignored". Indeed, there is correlation between solar activity and solar system motion (including the Sun), but it is the solar activity which influences on the Sun motion and thus far for this reason it influences on the motion of other planets in solar system. We should note also that in [16] authors came to conclusion that no evidence for planetary influence on solar activity, as well as in [17] authors concluded, by comparing various data of real observations, that accelerations of solar matter (at the tachocline level) may not be caused by planetary attractions (and so they concluded that the cause of the dynamo is purely solar). One additional source of reference is of importance in regard to the theme of our research, namely, it is work [18] where authors stated that they present evidence to show that changes in the Sun's equatorial rotation rate are synchronized with changes in its orbital motion about the barycentre of the solar system (but, nevertheless, they stated also that they are unable to suggest a plausible underlying physical cause for the coupling with respect to resonance between solar orbital motion and the meridional flow which takes place in the convective zone of the Sun).

## Appendix, A1 (comparison of estimations of gravitational forces, acting on Sun during its orbiting around barycenter of solar system).

Let us we estimate the gravitational forces, acting on Sun during its orbiting around barycenter of solar system:

Table 1. Comparison of estimations of gravitational forces, acting on Sun during its orbiting around barycenter of solar system.

| Masses of the planets (*solar system*), kg | Ratio $\left(\frac{m_{planet} / m_{Earth}}{M_{Sun} / m_{Earth}}\right) = \mu$ | Semimajor axis $a_p$ of the planet, AU | Possible (maximal) distance *between Sun-planet*, AU (10³ km) | Force, acting on Sun (maximal absolute value, in units of Jupiter force), $\left(\frac{m_{planet}}{M_{Jupiter}}\right)\left(\frac{a_{Jupiter}+0.01}{a_p+0.01}\right)^2$ |
|---|---|---|---|---|

| Mercury, $3.3\cdot10^{23}$ | $\left(\frac{0.055}{332{,}946}\right)=0.165\cdot10^{-6}$ | 0.387 AU | 0.39+0.01 AU | $\left(\frac{0.165}{954.51}\right)\left(\frac{5.22}{0.4}\right)^2\cong0.029$ |
|---|---|---|---|---|
| Venus, $4.87\cdot10^{24}$ | $\left(\frac{0.815}{332{,}946}\right)=2.448\cdot10^{-6}$ | 0.723 AU | 0.72+0.01 AU | $\left(\frac{2.448}{954.51}\right)\left(\frac{5.22}{0.73}\right)^2\cong0.131$ |
| Earth, $5.97\cdot10^{24}$ + Moon, $7.36\cdot10^{22}$ | $\left(\frac{1.0123}{332{,}946}\right)=3.040\cdot10^{-6}$ | 1 AU | 1+0.01 AU | $\left(\frac{3.040}{954.51}\right)\left(\frac{5.22}{1.01}\right)^2\cong0.085$ |
| Mars, $6.42\cdot10^{23}$ | $\left(\frac{0.107}{332{,}946}\right)=0.321\cdot10^{-6}$ | 1.523 AU | 1.52+0.01 AU | $\left(\frac{0.321}{954.51}\right)\left(\frac{5.22}{1.53}\right)^2\cong0.004$ |
| Jupiter, $1.9\cdot10^{27}$ | $\left(\frac{317.8}{332{,}946}\right)=954.509\cdot10^{-6}$ | 5.205 AU | 5.21+0.02 AU | 1 |

| Saturn, $5.69{\cdot}10^{26}$ | $\left(\frac{95.16}{332{,}946}\right) = 285.812 \cdot 10^{-6}$ | 9.582 | 9.58+0.01 AU | $\left(\frac{285.812}{954.51}\right)\left(\frac{5.22}{9.5}\right)^2 \cong 0.090$ |
|---|---|---|---|---|
| Uranus, $8.69{\cdot}10^{25}$ | $\left(\frac{14.37}{332{,}946}\right) = 43.160 \cdot 10^{-6}$ | 19.201 | 19.2+0.01 AU | $\left(\frac{43.16}{954.51}\right)\left(\frac{5.22}{19.21}\right)^2 \cong 0.003$ |
| Neptune, $1.02{\cdot}10^{26}$ | $\left(\frac{17.15}{332{,}946}\right) = 51.510 \cdot 10^{-6}$ | 30.048 | 30.05+0.01 AU | $\left(\frac{51.51}{954.51}\right)\left(\frac{5.22}{30.06}\right)^2 \cong 0.002$ |
| | | | | |
| Mass of object, kg | Ratio of masses (where Sun is considered as planet with respect to barycenter of solar system) | Variable semimajor axis $a_{p\ (Sun)}$ of the Sun, AU | Possible (maximal) distance *between Sun-barycenter*, AU ($10^3$ km) | Forces, acting on barycenter from Sun and *vice versa* (maximal absolute value, in units of Jupiter force above) |

| Sun, 1.02·$10^{30}$ | 1 | Max. 0.01 | 0.01 | $\left(\frac{M_{Sun}}{M_{Jupiter}}\right)\left(\frac{a_{Jupiter}+0.01}{a_{p(Sun)}}\right)^2 = \left(\frac{1020}{1.9}\right)\left(\frac{5.22}{0.01}\right)^2 \cong 1.463\cdot 10^8$ |
|---|---|---|---|---|
| Barycenter, ~ 1.02·$10^{30}$ | 1 + $10^{-6}$×(0.165+2.448+ 3.040+0.321+954.509 +285.812+43.160+ 51.510) = 1.001341 | Max. 0.01 | 0.01 | $\left(\frac{1021.4}{1.9}\right)\left(\frac{5.22}{0.01}\right)^2 \cong 1.465\cdot 10^8$ |

As we can see from Table 1, the difference between force attracting Sun to barycenter of solar system (its maximal absolute value, in units of Jupiter force above) relative to the force attracting barycenter to the Sun (in units of Jupiter force above) is at least in 200 thousand times more than the force acting from Jupiter on Sun in Sun's motion around barycenter of solar system (meanwhile, actions of other planets of solar system are obviously negligible in comparing with Jupiter, except Venus, Earth and Saturn, e.g. their combained contribution is less alltogether than 31%, Mercury's contribution is less than 3%, contributions of other planets is less than 1%, with respect to that one from Jupiter). It let us conclude that the dynamical cause that forcing the Sun to quasi-periodically orbit around the aforesaid barycenter is superior much more than even combined (possible) effect of gravitational attraction from all the planets of solar system together.

**Appendix, A2 (estimation for the required power of the series of explosion using nuclear weapon).**

Let us we estimate the required power of the series of explosion using nuclear weapon. As theoretical studies have shown, the radii of the zones of destruction and damage by the shock wave of nuclear and thermonuclear explosions of various powers are proportional to the cubic root of the ratio of TNT equivalents (where TNT equivalent is a convention for expressing energy, typically used to describe the energy released in an explosion):

$$\frac{R_2}{R_1} \cong \sqrt[3]{\frac{q_2}{q_1}}$$

here $R_2$ is the radius of the zone of destruction and damage by the shock wave of nuclear or thermonuclear explosion which should be evaluated ($q_2$ is the TNT equivalent of such the nuclear or thermonuclear explosion), $R_1$ is the known radius of the zone of destruction and damage by the shock wave of previous nuclear or thermonuclear explosion which has been done before with known results of such explosion ($q_2$ is its TNT equivalent).

Most known example of nuclear explosion in atmosphere was Big Ivan or The Tsar Bomba ("King of Bombs") [19-21], results of measurements and other data suggested the bomb yielded explosion approx. 58 Mt (in 1991, this estimation has been corrected by soviet scientists to the amount 50 Mt). Direct or indirect results (important for our estimation) were as follows:

- Consider the potential effects of a 50-megaton nuclear explosion, similar to the Tsar Bomba, detonated in October 1961, which shattered windows 780 kilometers away and sent a mushroom cloud rising above the stratosphere. The height of visible explosion (a mushroom cloud) was 67 km;

- Shock-wave 3 times has come cycling around the Earth (most fast in the first time, during 36.5 hours), so the longitudinal effect of nuclear explosion was elongated up to circa 3*{2*(3.1416)*6371 km} ≅ 120.1 thousand of kilometers;
- Atmosphere was ionised at distance few hundred kilometers, during 40 minutes.

Basing on the data above, we can conclude (as first approximation) that minimal distance of direct influence of mushroom cloud in atmospere (in case of Tsar Bomba explosion) was not less than 70 kilometers, maximal (most effective) distance of longitudinal effect of nuclear explosion appears to be 40 thousand of kilometers.

Meanwhile, depth of the convective solar zone is circa 150-200 thousand of kilometers, so we should use

$$\frac{q_2}{q_1} \cong \left(\frac{150}{120}\right)^3 \cong 1.95$$

not less than 2 the Tsar Bombs to achieve the effect of noticeable distortion of surface (or topology) of convective solar zone during not less than 36 hours (it is worth to note that explosion should be made at the middle-depth of convective solar zone). The aforementioned conclusion is valid if we would consider the spherical symmetry for the explosion process. In case of considering the explosion process to be similar to that one have been made in Earth atmosphere (in case of Tsar Bomba explosion) such explosion should be made at the depth of 100 kilometers, sufficiently close to the surface of convective solar zone. In this case, distortion of the surface of convective solar zone by the shock wave along with the mushroom cloud rising above the convective solar zone would influence on surface (or topology) of convective solar zone during sufficiently long-time period. Such influence will cause the local disbalance in self-resulting photo-gravitational force (driving Sun center motion relative to barycenter of Solar system) which will be caused by anisotropic radiating light and energy fluxes into space over all the surface of Sun during period of its full turnover, ~ circa 25 days.

The last but not least, we should especially note that the problem of transporting the nuclear bomb to the solar convective zone is a very hard engineering problem, not only due to obvious presence of hot plasma inside this zone (and also on a path of a spaceship to the target depth inside this zone, through the solar corona surrounding solar convective zone) which could be solved by multiple covering of spacecraft transporting nuclear bomb by special materials allowing the ablation process, but also for the reason that such natural restriction should be valid for the *motion* of the spacecraft in the vicinity of barycenter of Solar system [22] as the *Roche* limit for spacecraft's fly-by through the hot atmosphere and corona of Sun. To overcome the last restriction, spacecraft should be moving with sufficiently higher speed when fly-by closer than $4\ R_{Sun}$ through the solar corona surrounding solar convective zone (where $R_{sun}$ is the radius of Sun).

**Conflict of interest**

The authors declare that there is no conflict of interests regarding the publication of this article.

Remark regarding contributions of authors as below:

In this research, Dr. Sergey Ershkov is responsible for the main idea and general ansatz, suggested algorithm (how to move towards stars using Solar self-resulting photo-gravitational force), simple algebraic calculating, results of the article and also is responsible for the obtaining approximate estimations.

Prof. Dmytro Leshchenko is responsible for theoretical investigations and deep survey in literature on the problem under consideration.

All authors agreed with results and conclusions of each other in Sections 1-3.